\documentclass[aps,prx,reprint,superscriptaddress,footnote, onecolumn,notitlepage]{revtex4-1}

\usepackage{tikz}  
\usetikzlibrary{arrows,shapes,positioning,shadows,backgrounds,fit}
\usepackage{float}
\usepackage{relsize}
\usepackage{graphicx}
\usepackage{amsmath, bbm}
\usepackage{amssymb}
\usepackage{amsthm}
\usepackage{mathrsfs}
\usepackage{xcolor}
\usepackage{cancel}
\usepackage{bbold}
\usepackage{titlesec}
\usepackage{enumitem}  

\usepackage[normalem]{ulem}

\titlespacing{\section}{0ex}{2ex}{0.4ex}
\titleformat{name=\section}{\bf}{\thesection.}{.3em}{}

\usepackage{soul}
\usepackage{dsfont}

\def\be{\begin{eqnarray}}
\def\ee{\end{eqnarray}}
\newcommand{\tr}[1]{\text{Tr}\left(#1\right)}
\newcommand{\<}{\langle}
\renewcommand{\>}{\rangle}

\renewcommand{\ln}[1]{\mathrm{ln} \left({#1} \right)}

\theoremstyle{plain}

\definecolor{myblue}{rgb}{0.2,0.2,0.8}
\definecolor{myblack}{rgb}{0,0,0}
\definecolor{myurl}{rgb}{0.1,0.1,0.4}

\usepackage[colorlinks=true,citecolor=myblue,linkcolor=myblack,urlcolor=myurl]{hyperref}

\begin{document}

\title{Optimal control of dissipation and work fluctuations for rapidly driven systems} 

\author{Alberto Rolandi}
\affiliation{D\'{e}partement de Physique Appliqu\'{e}e,  Universit\'{e} de Gen\`{e}ve,  1211 Gen\`{e}ve,  Switzerland}

\author{Martí Perarnau-Llobet}
\affiliation{D\'{e}partement de Physique Appliqu\'{e}e,  Universit\'{e} de Gen\`{e}ve,  1211 Gen\`{e}ve,  Switzerland}

\author{Harry J.~D. Miller}
\affiliation{Department of Physics and Astronomy, The University of Manchester, Manchester M13 9PL, UK.}

\date{\today}

\begin{abstract}
To achieve efficient and reliable control of microscopic systems one should look for driving protocols that mitigate both the average dissipation and stochastic fluctuations in work. This is especially important in fast driving regimes in which the system is driven far out of equilibrium, potentially creating large amounts of unwanted entropy production. Here we characterise these optimal protocols in rapidly driven classical and quantum systems and prove that they consist of two discontinuous jumps in the full set of control variables. These jumps can be tuned to interpolate between processes with either minimal dissipation or minimal fluctuations, and in some situations allow for simultaneous minimisation. We illustrate our general results with rapidly driven closed quantum systems, classical bit erasure and a dissipative Ising chain driven close to a quantum phase transition.
\end{abstract}

\maketitle

Minimising dissipation is a central optimal control problem in stochastic and quantum thermodynamics \cite{Deffner}, and is especially important for ensuring efficient control of microscopic machines operating out of equilibrium. Careful consideration is needed when choosing control protocols that drive small systems from one state to another in finite-time, as generating too much dissipation leads to irreversibility \cite{Landi2021a} and thus hampers any thermodynamic performance. A key feature of microscopic systems, whether they be quantum or classical, is that they are also heavily influenced by stochastic fluctuations, hence thermodynamic quantities such as work and heat behave as random variables \cite{Seifert2012,Esposito2009}. Therefore from an optimal control perspective it is also desirable to ensure fluctuations around the mean dissipation are kept small in order to maintain precision and stability along a thermodynamic process. However, typically it is not possible to minimise the dissipation and fluctuations simultaneously and a compromise must be chosen. Current research aims to understand the interplay and unavoidable trade-offs between dissipation and fluctuations in classical-stochastic and quantum thermodynamic systems~\cite{Pietzonka2018,Solon2018,Horowitz2020,Holubec2022,miller2022finite}, and it remains an ongoing problem of how best to balance these two competing figures of merit in different scenarios.     

With regard to dissipation, or equivalently the average excess work done to drive a system from one state to another, optimal processes are well characterised in slowly driven or linear response regimes where methods from thermodynamic geometry can be employed \cite{Salamon1983,Zulkowski2012,Bonanca2014,Abiuso2020a,Brandner2020}. In this case a metric can be assigned to the control parameter space with minimum dissipation protocols achieved by driving the system along a geodesic path \cite{Sivak2012}. For classical stochastic systems operating in these close-to-equilibrium regimes the work fluctuation-dissipation relation holds \cite{Jarzynski1997d,Speck}, implying that paths of minimal dissipation simultaneously minimise the resulting work fluctuations. While this is not always satisfied by non-classical systems due to the impact of quantum coherence \cite{Miller2019}, an alternative geometric approach has been recently developed that can determine minimum fluctuation protocols for slowly driven quantum systems \cite{Miller2020,Mehboudi2022}. On the other hand, when operating much further from equilibrium over shorter timescales these geodesic paths are no longer relevant for minimising either of these quantities. It is of course desirable to understand how to optimise systems beyond linear response driving and where shorter operation times are needed. The central aim of this paper is to establish a general optimisation principle for minimising both the average excess work and its fluctuations in rapidly driven small systems. 

The study of driven Brownian particles first hinted at a key feature of minimum-dissipation protocols for fast driving; such protocols contain discontinuous jumps in the system control parameters or degrees of freedom~\cite{Schmiedl}. This has been further evidenced in a range of other systems through either analytic or numerical treatments of finite-time optimal control problems~\cite{Esposito2010,Muratore-ginanneschi2011,Geiger2010,Zulkowski2014a,Cavina}. More recently the optimality of these control parameter quenches has been proven \textit{in general} with regard to maximising the power and efficiency of microscopic heat engines with fast operation cycles \cite{Erdman2019,Cavina2020}, and furthermore proven optimal for minimising the average excess work done on classical stochastic systems rapidly driven from equilibrium~\cite{Blaber}. In contrast to dissipation, little is known about how to minimise work fluctuations under rapid driving, nor is it known how these protocols compare to those with minimal dissipation. In this paper we will prove that protocols with minimal fluctuations also consist of instantaneous jumps in the systems control parameters. Our result applies in full generality to any quantum or classical system whose generator is independent of the control parameter velocities. While sharing the same general design principle as minimal-dissipation protocols, these control variables typically need to jump to a distinct point in the parameter space, meaning that average excess work and work fluctuations cannot be simultaneously optimised.  Furthermore, and as we will illustrate, one practical advantage of this approach is that it enables us to optimise driving protocols for  complex many-body systems where exact results are lacking.

The paper is structured as follows; in Section~\ref{sec:expand} we derive general expressions for the average excess work and its variance for rapidly driven quantum systems, and in Section~\ref{sec:optimal} we present the general Euler-Lagrange equations for finding optimal protocols in this fast driving regime and show that all solutions consist of discrete jumps in the control parameter space. We then explore different scenarios where this optimisation scheme can be implemented; Section~\ref{sec:closed} focuses on closed quantum systems, whereas Section~\ref{sec:classical} concerns open quantum systems including erasure of a quantum dot and driving a classical and quantum Ising spin chain.

\section{Average excess work and fluctuations for fast driving}\label{sec:expand}

\

\noindent We will begin with a rather general treatment of a finite-dimensional quantum system subject to rapid time-dependent driving, which may be isolated or in contact with an environment. The Hamiltonian of the system is first parameterised by a set of $d$ scalar variables $\vec{\lambda}_t=\{\lambda_1 (t),\lambda_2 (t), ...\lambda_d(t)\}$ that can be controlled in time:
\begin{align}
    H(\vec{\lambda}_t)=H_0+\vec{\lambda}_t \cdot \vec{X}, \ \ \ \ \ \ t\in[0,\tau],
\end{align}
where $H_0$ denotes a fixed Hamiltonian in the absence of driving, and $\vec{X}$ are a set of corresponding conjugate observables with $\vec{X}=\{X_1,X_2, ...X_d\}$ which may be assumed time-independent without loss of generality. As we wish to compare the behaviour of different choices of driving protocols $\gamma:t\to \vec{\lambda}_t$, it will be useful to define the set of protocols $\gamma\in\mathscr{C}$ that have a fixed initial and final value:
\begin{align}\label{eq:control}
    \gamma:t\to \vec{\lambda}_t\in \mathscr{C}=\{ \vec{\lambda}_t\in\mathbb{R}^d \ | \ \vec{\lambda}_0=\vec{\lambda}_A,  \ \ \  \vec{\lambda}_\tau=\vec{\lambda}_B  \} .
\end{align}
 For now we can assume the evolution is given by a Markovian generator of the form
\begin{align}\label{eq:gen}
    \dot{\rho}(t)=\mathcal{L}_{\vec{\lambda}_t}[\rho(t)]; \ \ \ \ \rho(0)=\pi(\vec{\lambda}_A).
\end{align}
with a thermal initial condition, where
\begin{align}
    \pi(\vec{\lambda})=\frac{e^{-\beta H(\vec{\lambda})}}{Z(\vec{\lambda})}; \ \ \ \ \ \  Z(\vec{\lambda}):=\tr{e^{-\beta H(\vec{\lambda})}}
\end{align}
is the corresponding Gibbs state at inverse temperature $\beta$. The most notable part of this assumption is that the generator is \textit{independent of the velocity $d\vec{\lambda}_t/dt$}, and depends only on the local values of $\vec{\lambda}_t$. This is readily satisfied by isolated quantum systems evolving unitarily, adiabatically driven open quantum systems \cite{Albash2012} and Markovian dynamics for classical/quasi-classical systems driven by scalar potentials. On the other hand, open quantum quantum systems driven non-adiabatically may not meet this requirement \cite{Dann}.  

The control protocol will result in some work done on the system since it is driven out of equilibrium. Due to both thermal and quantum fluctuations, the work $W$ is a stochastic quantity and its statistics are described by a distribution $P(W)$. In a closed, unitarily-driven system this distribution can be ascertained from projective measurements the system's Hamiltonian at the beginning and end of the process \cite{Talkner2007b,Esposito2009,Roncaglia2014c}. For weakly-coupled open systems one can similarly determine the work distribution by monitoring a set of quantum jumps as the system exchanges energy with its environment \cite{Horowitz2013b,Elouard2018}. In any case, the average work $\langle W \rangle$ and its variance $\sigma_W^2=\langle W^2 \rangle-\langle W \rangle^2$ are given by the following general form \cite{Suomela2014,Miller2019}:
\begin{align}\label{eq:av_work}
    &\langle W \rangle:=\int^{\tau}_0 dt \ \frac{d\vec{\lambda}_t^T}{dt} \  \tr{\vec{X}\rho(t)}, \\
    &\sigma_W^2:=2 \ \mathbb{R}e\int^\tau_0 dt \int^t_0 dt' \ \tr{\dot{H}(\vec{\lambda}_t)\overleftarrow{P}(t,t')\big[\Delta_{\rho_{t'}}\dot{H}(\vec{\lambda}_{t'}) \rho_{t'}\big]},\label{eq:var_work}
\end{align}
where  we denote $\Delta_\rho A=A-\tr{A\rho}$ and
\begin{align}
    \overleftarrow{P}(t,t')[(.)]:=\overleftarrow{\mathcal{T}}\text{exp}\bigg(\int^{t}_{t'} d\nu \ \mathcal{L}_{\vec{\lambda}_\nu}\bigg)[(.)],
\end{align}
is the time-ordered propagator. As a quantifier for the degree of irreversibility associated with the process, the average \textit{excess} work done on the system is defined as
\begin{align}
    \langle W_{\text{ex}}\rangle&=\langle W \rangle-\Delta F, 
\end{align}
where $\Delta F=F(\vec{\lambda}_B)-F(\vec{\lambda}_A)$ is the change in equilibrium free energy, $F(\vec{\lambda}):=-\beta^{-1}\log Z(\vec{\lambda})$. The excess work disappears $\langle W_{\text{ex}} \rangle \rightarrow 0$ in  quasistatic processes  where the system is always in thermal equilibrium, which also implies absence of work fluctuations  due to the work fluctuation-dissipation relation $\beta \sigma_W^2/2 = \langle W_{\text{ex}}\rangle$ holding valid in this limit~\cite{Jarzynski1997d,Speck}. For  non-equilibrium processes, both  $\langle W_{\text{ex}} \rangle $ and $\sigma_W$ will become relevant, and we expect their magnitudes to increase with the speed of the process (i.e. as $\tau$ decreases).  
 Our goal is then to investigate which protocols  $\gamma$ in \eqref{eq:control}  give the smallest values of average dissipation $\langle W_{\text{ex}}\rangle$ and work fluctuations $\sigma_W$ respectively.  In general, computing and optimising the work moments relies on knowing an exact solution to the dynamics~\eqref{eq:gen}. While this is not generally tractable,  we will demonstrate that this control problem becomes considerably simpler in fast driving regimes (i.e. when the overall time $\tau$ taken to go from $\vec{\lambda}_A$ to $\vec{\lambda}_B$ is small relative to the characteristic timescales of the system).

We first quantify precisely what we mean by a rapid protocol by defining a characteristic timescale $\tau_{\text{c}}$ for the generator given by \cite{Cavina2020}
\begin{align}\label{eq:timescale}
    \tau_{\text{c}}^{-1}=\max_{\vec{\lambda}_t\in \mathscr{C}}||\mathcal{L}_{\vec{\lambda}_t}||,
\end{align}
where we introduce a norm
\begin{align}
    ||\mathcal{L}_{\vec{\lambda}}||=\max_{\tr{O}< \infty}\frac{||\mathcal{L}_{\vec{\lambda}_t}[O]||_1}{||O||_1}
\end{align}
and $||A||_1=\tr{\sqrt{A^\dagger A}}$. For a finite-dimensional unitary generator, this parameter is bounded by the operator norm of the Hamiltonian, while for systems undergoing non-unitary dynamics with a unique fixed point then $\tau_{\text{c}}$ bounds the shortest relaxation timescale associated with the system. Overall, this gives us a definition of the fast driving regime which assumes that the total duration is short enough such that $\tau\ll \tau_c$. To see how this approximation impacts the work moments, let us rewrite the evolution in dimensionless time:
\begin{align}\label{eq:gen2}
    \dot{\tilde{\rho}}(s)=\tau\mathcal{L}_{\vec{\lambda}_s}[\tilde{\rho}(s)]; \ \ \ \ \tilde{\rho}(0)=\pi(\vec{\lambda}_A), \ \ \ \ s=\frac{t}{\tau}\in[0,1]
\end{align}
In these units the work done is 
\begin{align}
    \langle W \rangle:=\int^{1}_0 dt \ \frac{d\vec{\lambda}_s^T}{ds} \  \tr{\vec{X} \ \tilde{\rho}(s)},
\end{align}
We can expand the solution to~\eqref{eq:gen2} as a Dyson series:
\begin{align}\label{eq:dyson}
    \tilde{\rho}(s)=\pi(\vec{\lambda}_A)+\sum^{\infty}_{n=1}\tau^n\int^s_0 dt_{n} \int^{t_{n}}_0 dt_{n-1}...\int^{t_2}_{0}dt_1 \  \mathcal{L}_{\vec{\lambda}_{t_{n}}}\mathcal{L}_{\vec{\lambda}_{t_{n-1}}}...\mathcal{L}_{\vec{\lambda}_{t_1}}[\pi(\vec{\lambda}_A)].
\end{align}
where $0\leq t_1\leq ...\leq t_n\leq s$. Consider the first two terms in the expansion, 
\begin{align}
\sigma(s)=\pi(\vec{\lambda}_A)+\tau\int^s_0 dt' \mathcal{L}_{\vec{\lambda}_{t'}}[\pi(\vec{\lambda}_A)],
\end{align}
From our definition of the characteristic timescale~\eqref{eq:timescale} we have
\begin{align}\label{eq:approx}
    ||\tilde{\rho}(s)-\sigma(s)||_1\leq\mathcal{O}([\tau/\tau_c]^2). 
\end{align}
Therefore we can approximate the state by $\tilde{\rho}(s)\simeq \sigma(s)$ so long as $\tau\ll \tau_c$. This approximation should be contrasted with the slow driving regime that treats the opposite case, $\tau_c\ll \tau$. Assuming the dynamics has a thermal fixed point, a slow driving approximation implies that the system stays close to equilibrium at all times, ie. $\tilde{\rho}(s)\simeq \pi(\vec{\lambda}_s)+\delta \rho(\vec{\lambda}_s) $ with $\delta \rho(\vec{\lambda}_s)$ a linear correction to the instantaneous thermal state \cite{Cavina2017}. In the present context, the Dyson series allows us to perform the inverse of this expansion, with rapid driving moving the system far from instantaneous equilibrium instead. We can use~\eqref{eq:approx} to now approximate the excess work, yielding
\begin{align}\label{eq:av_expand}
    \nonumber \langle W_{\text{ex}}\rangle&\simeq k_B T S(\pi(\vec{\lambda}_A)||\pi(\vec{\lambda}_B))+\tau\int^1_0 \frac{d\vec{\lambda}_s^T}{ds}\int^s_0 dt' \ \tr{\vec{X} \ \mathcal{L}_{\vec{\lambda}_{t'}}[\pi(\vec{\lambda}_A)]}, \\
    &=k_B T S(\pi(\vec{\lambda}_A)||\pi(\vec{\lambda}_B))+\int^\tau_0 dt \ \frac{d\vec{\lambda}_t^T}{dt}\int^t_0 dt' \ \vec{R}(\vec{\lambda}_{t'}).
\end{align}
where we define the relative entropy 
\begin{align}
    S(\rho_1||\rho_2)=\tr{\rho_1 \log \rho_1}-\tr{\rho_1 \log \rho_2}
\end{align}
and the quantum \textit{initial force relaxation rate} (IFRR):
\begin{align}
    \vec{R}(\vec{\lambda}_t):=\big\<\mathcal{L}^\dagger_{\vec{\lambda}_{t}}[\vec{X}]\big\>_{A}
\end{align}
where  $\langle (.) \rangle_{A}=\tr{(.)\pi(\vec{\lambda}_A)}$ is an average with respect to the initial equilibrium state. We then do an integration by parts to obtain
\begin{align}\label{eq:fastwork}
    \langle W_{\text{ex}}\rangle= k_B T S\big(\pi(\vec{\lambda}_A)||\pi(\vec{\lambda}_B)\big) + \int^\tau_0 dt \ \big[\vec{\lambda}_B - \vec{\lambda}_t\big]^T \vec{R}(\vec{\lambda}_t).
\end{align}
The first term represents the excess work from a perfect Hamiltonian quench \cite{Scandi2019}, while the second term gives the leading order correction for a protocol at finite speed. This expansion agrees with the results of \cite{Blaber} for classical Focker-Planck dynamics, now generalised to a fully quantum regime. 

Turning attention now to the work fluctuations, we convert the expression~\eqref{eq:var_work} into dimensionless units, so  
\begin{align}
    \sigma_W^2=2 \ \mathbb{R}e\int^1_0 ds \int^s_0 ds' \ \tr{\dot{H}(\vec{\lambda}_s)\overleftarrow{\mathcal{T}}\text{exp}\bigg(\tau\int^{s}_{s'} d\nu \ \mathcal{L}_{\vec{\lambda}_\nu}\bigg)\big[\Delta_{\tilde{\rho}(s')}\dot{H}(\vec{\lambda}_{s'}) \tilde{\rho}(s')\big]}, 
\end{align}
where we can approximate the propagator using the Dyson series again, so that
\begin{align}
    \nonumber\sigma_W^2&\simeq   2 \ \mathbb{R}e\int^1_0 ds \int^s_0 ds' \ \tr{\dot{H}(\vec{\lambda}_s) \ \Delta_{\tilde{\rho}(s')}\dot{H}(\vec{\lambda}_{s'}) \tilde{\rho}(s')} \\
    \nonumber& \ \ \ \ \ \ \ \ \ \ \ \ \ \ \ \ \  \ \ \ \ \ +2\tau \ \mathbb{R}e\int^1_0 ds \int^s_0 ds' \int^{s}_{s'} d\nu \ \tr{ \dot{H}(\vec{\lambda}_s)\mathcal{L}_{\vec{\lambda}_\nu}\big[\Delta_{\pi(\vec{\lambda}_A)}\dot{H}(\vec{\lambda}_{s'}) \pi(\vec{\lambda}_A)\big]}, 
\end{align}
Applying the Dyson expansion to a shifted observable yields
\begin{align}
    \Delta_{\tilde{\rho}(s)}A \simeq \Delta_{\tilde{\sigma}(s)}A = \Delta_{\pi(\vec{\lambda}_A)}A - \tau\int_0^s ds'~\tr{ A\ \mathcal{L}_{\vec{\lambda}_{s'}}[\pi(\vec{\lambda}_A)]}
\end{align}
Using this expansion, the Dyson expansion of the state and converting back to original units of time we get
\begin{align}\label{eq:var_expand}
 \nonumber \sigma_{W}^2 \simeq k_B^2 T^2 V\big(\pi(\vec{\lambda}_A)||\pi(\vec{\lambda}_B)\big) &- 2 \ \mathbb{R}e\int^\tau_0 dt \int^t_0 dt'\int^{t'}_0 d\nu \ \tr{\dot{H}(\vec{\lambda}_t)\pi(\vec{\lambda}_A)} \tr{\dot{H}(\vec{\lambda}_{t'})\mathcal{L}_{\vec{\lambda}_\nu}[\pi(\vec{\lambda}_A)]} , \\
 \nonumber& + 2 \ \mathbb{R}e\int^\tau_0 dt \int^t_0 dt'\int^{t'}_0 d\nu \ \tr{\dot{H}(\vec{\lambda}_t) \ \Delta_{\pi(\vec{\lambda}_A)}\dot{H}(\vec{\lambda}_{t'})\mathcal{L}_{\vec{\lambda}_\nu}[\pi(\vec{\lambda}_A)]} \\
 & +2 \ \mathbb{R}e\int^\tau_0 dt \int^t_0 dt' \int^{t}_{t'} d\nu \ \tr{ \dot{H}(\vec{\lambda}_t)\mathcal{L}_{\vec{\lambda}_\nu}\big[\Delta_{\pi(\vec{\lambda}_A)}\dot{H}(\vec{\lambda}_{t'}) \ \pi(\vec{\lambda}_A)\big]},
\end{align}
where the first term is the \textit{relative entropy variance} \cite{Guarnieri2018a}:
\begin{align}
    V(\rho_1||\rho_2)=\tr{\rho_1 (\log \rho_1-\log \rho_2)^2}-S^2(\rho_1||\rho_2).
\end{align}
Let us first define the initial force correlation matrix $\mathbf{G}(\vec{\lambda})$, with elements
\begin{align}
    \big[\mathbf{G}(\vec{\lambda})\big]_{jk}:=\frac{1}{2} \big\<\mathcal{L}^\dagger_{\vec{\lambda}}\big[\{\Delta X_j,\Delta X_k\}\big]\big\>_{A},
\end{align}
where $\{X,Y\}=XY+YX$ is the anti-commutator and we define shifted force observables
\begin{align}
    \Delta X_j:=X_j-\big<X_j\big\>_{A}.
\end{align}
We also need to introduce another correlation function given by
\begin{align}
    \big[\mathbf{B}(\vec{\lambda})\big]_{jk}:=\big<\{\mathcal{L}^\dagger_{\vec{\lambda}_t}[\Delta X_j],\Delta X_k\}\big\>_A.
\end{align}
Then the nested integrals in~\eqref{eq:var_expand} can be converted into a single one using integration by parts twice:
\begin{align}\label{eq:fast_var}
 \sigma_{W}^2 = k_B^2 T^2 V\big(\pi(\vec{\lambda}_A)||\pi(\vec{\lambda}_B)\big)
 +\int^\tau_0 dt \ \big[\vec{\lambda}_B-\vec{\lambda}_t\big]^T \mathbf{G}(\vec{\lambda}_t)\big[\vec{\lambda}_B-\vec{\lambda}_t\big]+ \big[\vec{\lambda}_B-\vec{\lambda}_t\big]^T \mathbf{B}(\vec{\lambda}_t) \big[\vec{\lambda}_t-\vec{\lambda}_A\big].
\end{align}
As we saw with the average excess work, the first term here is what one would expect for work fluctuations via an instantaneous quench \cite{Scandi2019}, while the two terms in the integral are leading order corrections for a finite speed protocol. The expressions~\eqref{eq:fastwork} and~\eqref{eq:fast_var} are our first main results, and will now form the basis for finding optimal protocols in the fast driving regime.  

\section{Optimality of instantaneous jump protocols}\label{sec:optimal}

\

Our aim is now to determine control protocols~\eqref{eq:control} that minimise the average excess work and the work fluctuations. For this it is useful to define the \textit{short-term power savings} \cite{Blaber}, defined as
\begin{align}\label{eq:power}
    P_{\text{save}}:=\tau^{-1}\bigg(k_B T S\big(\pi(\vec{\lambda}_A)||\pi(\vec{\lambda}_B)\big)-\langle W_{\text{ex}}\rangle\bigg),
\end{align}
which quantifies any additional reduction to the rate of work done provided by the finite-time protocol beyond that of an instantaneous quench. In a similar fashion we also introduce the \textit{short-term constancy savings},
\begin{align}\label{eq:const}
    C_{\text{save}}:=\tau^{-1}\bigg(k_B^2 T^2 V\big(\pi(\vec{\lambda}_A)||\pi(\vec{\lambda}_B)\big)-\sigma_{W}^2\bigg),
\end{align}
This measures the reductions to the rate of work fluctuations from a short-time protocol. These are now the two objectives to maximise in our control problem. Using our short-time approximations to both the average excess work~\eqref{eq:fastwork} and work fluctuations~\eqref{eq:fast_var}, a general optimisation principle becomes immediately apparent for this regime. Since the integrands appearing in~\eqref{eq:fastwork} and~\eqref{eq:fast_var} are each independent of the control velocity $d\vec{\lambda}/dt$, we can infer that optimal protocols will consist of an instantaneous jump from $\vec{\lambda}_A$ to a point in the parameter space, remaining there for the total duration $\tau$ and concluding with another instantaneous jump to the final boundary point $\vec{\lambda}_B$. We will denote the control values that maximise $P_{\text{save}}$ and $C_{\text{save}}$ respectively by $\vec{\zeta}$ and $\vec{\Lambda}$, which are determined by the solutions to the following \textit{distinct} Euler-Lagrange equations:
\begin{align}\label{eq:EL_av}
    \vec{R}(\vec{\zeta})=\nabla_{\vec{\lambda}}\bigg(\big[\vec{\lambda}_B - \vec{\zeta}\big]^T\vec{R}(\vec{\lambda})\bigg)\bigg|_{\vec{\lambda}=\vec{\zeta}} 
\end{align}
and
\begin{align}\label{eq:EL_var}
    \nabla_{\vec{\lambda}}\bigg(\big[\vec{\lambda}_B-\vec{\lambda}\big]^T \mathbf{G}(\vec{\lambda})\big[\vec{\lambda}_B-\vec{\lambda}\big]\bigg)\bigg|_{\vec{\lambda}=\vec{\Lambda}}= \nabla_{\vec{\lambda}}\bigg(\big[\vec{\lambda}-\vec{\lambda}_B\big]^T \mathbf{B}(\vec{\lambda}) \big[\vec{\lambda}-\vec{\lambda}_A\big]\bigg)\bigg|_{\vec{\lambda}=\vec{\Lambda}}.
\end{align}
The maximal short-term power savings are then given by
\begin{align}\label{eq:max_p}
    P_{\text{save}}\leq P^{*}_{\text{save}}:=\big[\vec{\zeta}-\vec{\lambda}_B\big]^T \vec{R}(\vec{\zeta}),
\end{align}
which is saturated via the jump protocol
\begin{align}
    \vec{\lambda}_t=\vec{\lambda}_A+[\vec{\zeta}-\vec{\lambda}_A]\theta(t)+[\vec{\lambda}_B-\vec{\zeta}]\theta(t-\tau),
\end{align}
where $\theta(t)$ denotes the Heaviside step function. The optimality of such processes was proven in \cite{Blaber} for classical systems. We have here shown that the same result applies to quantum mechanical systems, provided that the dynamical generator~\eqref{eq:gen} remains independent of $d\vec{\lambda}/dt$. As a more significant result, we can now see that it is possible to reduce fluctuations using a similar instantaneous jump protocol, albeit with a different choice of point in the parameter space. The maximum short-term constancy savings are given by
\begin{align}\label{eq:max_c}
    C_{\text{save}}\leq C^{*}_{\text{save}}:= \big[\vec{\Lambda}-\vec{\lambda}_B\big]^T\bigg( \mathbf{G}(\vec{\Lambda})\big[\vec{\lambda}_B-\vec{\Lambda}\big]+  \mathbf{B}(\vec{\Lambda}) \big[\vec{\Lambda}-\vec{\lambda}_A\big]\bigg).
\end{align}
which is saturated by jumping to $\vec{\Lambda}$ instead, so that
\begin{align}
    \vec{\lambda}_t=\vec{\lambda}_A+[\vec{\Lambda}-\vec{\lambda}_A]\theta(t)+[\vec{\lambda}_B-\vec{\Lambda}]\theta(t-\tau).
\end{align}
In general the values of $\vec{\zeta}$ and $\vec{\Lambda}$ will not typically coincide, implying a trade-off between minimised excess work versus minimal fluctuations. This can remain the case even in quasi-classical regimes where only changes to the energy levels of the system are allowed, as well as fully classical systems that admit a phase space description. This should be contrasted with slow driving or linear response regimes, which allow for simultaneous optimisation of the average and variance due to the validity of the fluctuation dissipation relation in the absence of quantum friction \cite{Speck,Miller2019}. However, depending on the particular Hamiltonian parameters and dynamics it is still possible to find situations where $\vec{\zeta}=\vec{\Lambda}$ and simultaneous optimisation is possible, as we will highlight in subsequent sections. 

Before we proceed it is important to highlight some consistency requirements needed to implement a jump protocol. As we are restricted to operating in fast driving regimes, this places restrictions on the set of points one can jump to in order to ensure that the Taylor expansions used in~\eqref{eq:av_expand} and~\eqref{eq:var_expand} remain valid. In the Appendix it is shown that errors associated with these approximations scale according to
\begin{align}\label{eq:errorbound}
    |\langle W_{\text{ex}}\rangle^{\text{true}}-\langle W_{\text{ex}}\rangle^*|\leq \Delta h(\vec{\xi})  \ \mathcal{O}(\tau^2/\tau_c^2),
\end{align}
\begin{align}
    |(\sigma_W^2)^{\text{true}}-(\sigma^2_W)^*|\leq \Delta h^2(\vec{\Lambda})  \ \mathcal{O}(\tau^2/\tau_c^2),
\end{align}
where $\langle W_{\text{ex}}\rangle^{\text{true}}$ and $(\sigma_W^2)^{\text{true}}$ denote the exact values of~\eqref{eq:av_work} and~\eqref{eq:var_work} with respect to the jump protocols, while $\langle W_{\text{ex}}\rangle^*$ and $(\sigma^2_W)^*$ are the corresponding values with respect to the fast driving approximations~\eqref{eq:fastwork} and~\eqref{eq:fast_var}. Furthermore, we introduce the Hamiltonian magnitude
\begin{align}
    \Delta h(\vec{\lambda}):=2\max\big\{||H(\vec{\lambda})-H(\vec{\lambda}_A)||_1,||H(\vec{\lambda}_B)-H(\vec{\xi})||_1\big\}.
\end{align}
This tells us that one cannot jump arbitrarily far from the boundary points $\vec{\lambda}_A,\vec{\lambda}_B$ as this would lead to a large $\Delta h$ and hence invalidate the fast driving approximation. Therefore any freedom in setting the magnitude of $\vec{\xi}$ and $\vec{\Lambda}$ must take these bounds into account, discounting arbitrarily large values of both $P^{*}_{\text{save}}$ and $C^{*}_{\text{save}}$. For the remainder of the paper we now demonstrate the utility of these jump protocols in a range of different types of system.

\section{Closed quantum systems}\label{sec:closed}
\

\noindent As a starting point we consider an isolated quantum system whose dynamics are given by the time-dependent Liouville-von Neumann equation:
\begin{align}
\label{eq:evclosed}
    \mathcal{L}_{\vec{\lambda}_t}[ (\cdot)   ]=-\frac{i}{\hbar}\big[H(\vec{\lambda}_t),(\cdot)\big].
\end{align}
The work statistics of quenched isolated systems are well studied, particularly in the context of many-body quantum systems~\cite{Fusco2014a,Arrais2019}. Our formalism can now be used to calculate the leading short-time corrections to the excess work and fluctuations arising when the (instantaneous) quenches are replaced by fast Hamiltonian ramps, and then subsequently minimise them using the appropriate jump protocols outlined in the previous section. For closed, finite dimensional systems it is clear that the characteristic time scale is $\tau_c \sim\hbar/E_{\max}(\vec{\lambda})$, where $E_{\max}(\vec{\lambda})$ denotes the maximum energy eigenstate of $H(\vec{\lambda})$, and we set $\tau\ll \tau_c$ to establish the fast driving regime. The relevant initial force relaxation rate and correlation functions are found to be
\begin{align}
    \vec R(\vec\lambda) &= -\frac{i}{\hbar}\big\<\big[\vec X,H(\vec\lambda)\big]\big\>_A~,\\
    \big[\mathbf{G}(\vec{\lambda})\big]_{jk} &= -\frac{i}{2\hbar}\big\<\big[\big\{\Delta X_j,\Delta X_k\big\},H(\vec\lambda)\big]\big\>_A~,\\
    \big[\mathbf{B}(\vec{\lambda})\big]_{jk} &= -\frac{i}{\hbar}\big\<\big\{\big[\Delta X_j,H(\vec\lambda)\big],\Delta X_k\big\}\big\>_A~.
\end{align}
The  short-time power savings are then
\begin{align}
    P_{\text{save}}:=\frac{i}{\tau\hbar}\int^\tau_0 dt \  \big\<\big[H(\vec\lambda_B),H(\vec\lambda_t)\big]\big\>_A \label{eq:P_unitary}
\end{align}
while the constancy savings are
\begin{align}
    \nonumber C_{\text{save}}:=\frac{i}{\tau\hbar}\int^\tau_0 dt \ \Big(
    \big\<\big[H(\vec\lambda_B)^2,H(\vec\lambda_t)\big]\big\>_A &- \big\<\big\{H(\vec\lambda_A),\big[H(\vec\lambda_B),H(\vec\lambda_t)\big]\big\}\big\>_A \\ 
     &- 2\big\<H(\vec\lambda_B)-H(\vec\lambda_A)\big\>_A\big\<\big[H(\vec\lambda_B),H(\vec\lambda_t)\big]\big\>_A
    \Big) \label{eq:C_unitary}
\end{align}
We can already see from \eqref{eq:P_unitary} and \eqref{eq:C_unitary} that if $H(\vec\lambda_B)$ and $H(\vec\lambda_A)$ commute, or $\vec\lambda_t$ is chosen such that $H(\vec\lambda_t)$ commutes with either $H(\vec\lambda_B)$ or $H(\vec\lambda_A)$, then the integrand is exactly $0$ - which directly follows using the cyclic property of the trace. Therefore if $\vec\lambda_t$ is a linear combination of $\vec\lambda_A$ and $\vec\lambda_B$ the first order correction vanishes. An immediate consequence is that a naive protocol that linearly interpolates between the initial and final Hamiltonian in a closed system is equivalent to a quench up to first order in driving speed. Similarly, it follows directly (and without approximation) from \eqref{eq:evclosed} that if $H(\vec\lambda_B)$, $H(\vec\lambda_A)$ and $H(\vec\lambda_t)$ commute for all $t$ then the state does not evolve in time and therefore the protocol is equivalent to a quench. But we can note that here we get the same result (up to first order) with a weaker assumption on the commutation relations between the Hamiltonians.

We now choose a jump protocol to maximise either variable, and a simple argument can be used to determine the optimal points $\vec{\xi}$ and $\vec{\Lambda}$. First notice that both $P_{\text{save}}$ and $C_{\text{save}}$ are linear in the control variables, so that their gradients are independent of $\vec{\lambda}$:
\begin{align}\label{eq:gradP}
    \nabla_{\vec\lambda} P_{\text{save}} =  \frac{i}{\hbar} \big\<\big[H(\vec\lambda_B),\vec X\big]\big\>_A,
\end{align}    
\begin{align}\label{eq:gradC}     
    \nabla_{\vec\lambda} C_{\text{save}} =  \frac{i}{\hbar}\Big(
    \big\<\big[H(\vec\lambda_B)^2,\vec X\big]\big\>_A - \big\<\big\{H(\vec\lambda_A),\big[H(\vec\lambda_B),\vec X\big]\big\}\big\>_A - 2\big\<H(\vec\lambda_B)-H(\vec\lambda_A)\big\>_A\big\<\big[H(\vec\lambda_B),\vec X\big]\big\>_A
    \Big)~. 
\end{align}
The fact that these gradients are independent of $\vec\lambda$ implies that the optimal points $\vec{\xi}$ and $\vec{\Lambda}$ are vectors pointing in the direction of the respective gradients, with the norm chosen as large as possible. However, as argued in the previous section there is a limitation to how big this norm can be: the larger this norm is chosen the larger the error of the approximation is. In particular, setting $|\vec{\xi}|\gg |\vec\lambda_A |, |\vec\lambda_B |$ gives $|\vec{\xi}|\propto \Delta h(\vec{\xi})$ while $\tau_c \propto 1/|\vec{\xi}|$. Comparing with our error bound~\eqref{eq:errorbound} we see that in this case
\begin{align}\label{eq:error_closed}
    |\langle W_{\text{ex}}\rangle^{\text{true}}-\langle W_{\text{ex}}\rangle^*|\leq \mathcal{O}(|\vec{\xi}|^3\tau^2),
\end{align}
which clearly limits how large the norm can be chosen relative to the duration of the protocol. A similar argument applies to the constancy savings and norm of the optimal point $\vec{\Lambda}$. 

We can make some further inferences about the relation between the different jumps $\vec{\xi}$ and $\vec{\Lambda}$. It follows from the commutators in~\eqref{eq:gradP} and~\eqref{eq:gradC} that
\begin{align}\label{eq:grads_unitary}
    \vec{\lambda}_A\cdot \nabla_{\vec\lambda} P_{\text{save}}=\vec{\lambda}_A\cdot \nabla_{\vec\lambda} C_{\text{save}}=\vec{\lambda}_B\cdot \nabla_{\vec\lambda} P_{\text{save}}=\vec{\lambda}_B\cdot \nabla_{\vec\lambda} C_{\text{save}}=0,
\end{align}
which means that both gradients are orthogonal to $\vec\lambda_A$ and $\vec\lambda_B$. As was said before, this implies that if the protocol consists of a linear combination of $H_A$ and $H_B$ then the correction will be zero. But if the Hamiltonian has $d\leq 2$ controllable parameters it is impossible for $\vec\lambda_t$ to be linearly independent from $\vec\lambda_A$ and $\vec\lambda_B$. Therefore, regardless of the type of driving, with $d\leq 2$ controllable parameters the correction is always zero.

It is interesting to consider what happens when we can control exactly three parameters, $d=3$. Eq.~\eqref{eq:grads_unitary} constrains the gradients of~\eqref{eq:gradP} and~\eqref{eq:gradC} to be parallel, which implies a direct relation between the variation of power and fluctuations
\begin{align}
    dP_{\text{save}} =\pm \frac{||\nabla_{\vec\lambda} P_{\text{save}}  ||_{\vec{\lambda}=\vec{\xi}}}{||\nabla_{\vec\lambda} C_{\text{save}}  ||_{\vec{\lambda}=\vec{\Lambda}}} \ dC_{\text{save}}~,
\end{align}
where the sign is positive if the gradients are oriented in the same direction and negative otherwise. If the sign is positive we can optimize fluctuations and excess work simultaneously with $\vec{\xi}=\vec{\Lambda}$, if the sign is negative we have a direct trade-off between savings in power and constancy.

\subsection{Driven qubit}
\begin{figure}[H]
	\centering
	\includegraphics[width=.9\textwidth]{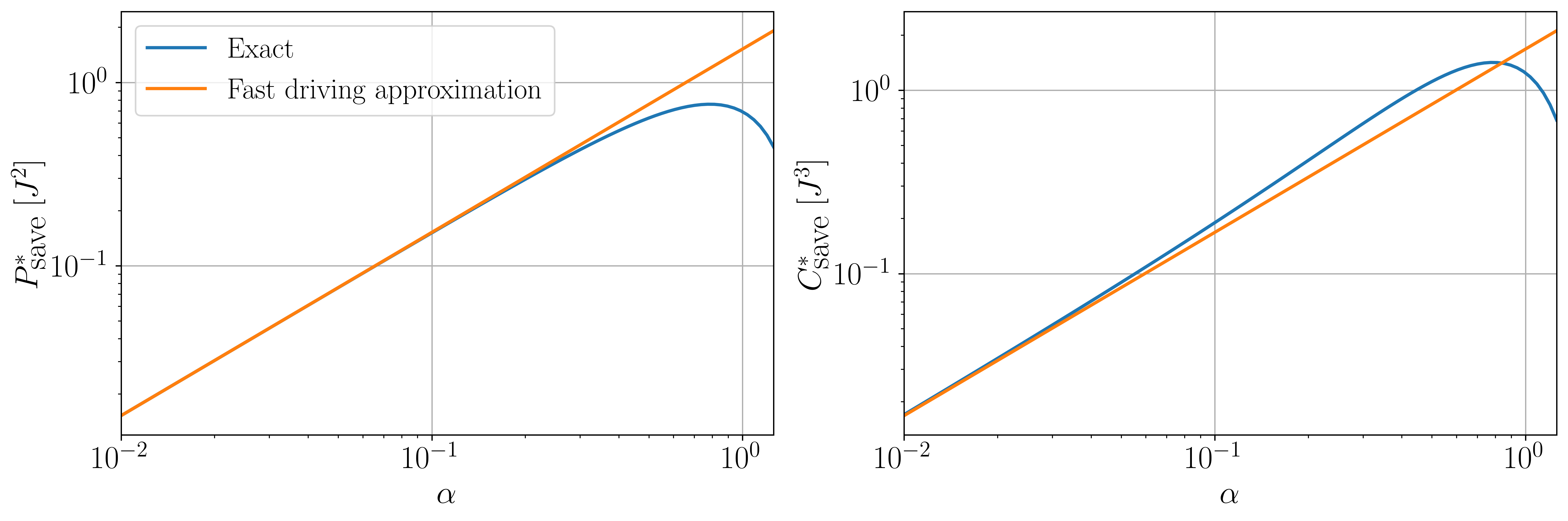}
	\caption{Comparison of $P_{\text{save}}^*$ and $C_{\text{save}}^*$ (in units of $J^2$ and $J^3$ respectively) in the exact case and fast driving approximation as a function of $\alpha$ for a jump protocol with $H_A = J\sigma^x$, $H_B = J\sigma^z$ and $J\tau = \beta J= 1$.}
	\label{fig:qubit}
\end{figure}

As an illustrative example we can compute and optimize the excess work and work fluctuations of a qubit that is undergoing unitary evolution. The most general Hamiltonian for a qubit is
\begin{equation}
    H(\vec{\lambda}) = J \vec\lambda \cdot \vec\sigma~,
\end{equation}
where $\vec\sigma = (\sigma^x,\sigma^y,\sigma^z)$ is the Pauli vector, $J$ is an energy scale and $\vec\lambda = (\lambda^x,\lambda^y,\lambda^z)$ are dimensionless parameters that characterize the Hamiltonian. Using the property that $(\vec a \cdot \vec\sigma) (\vec b \cdot \vec\sigma) = (\vec a \cdot \vec b) \mathbb{1} + i(\vec a \wedge \vec b)\cdot \vec\sigma$ we can find that the thermal state can be written in the following way
\begin{equation}
    \pi(\vec{\lambda})=\frac{e^{-\beta H(\vec{\lambda})}}{Z(\vec{\lambda})} = \frac{1}{2}\mathbb{1} - \frac{1}{2}\tanh(\beta J \|\vec\lambda\|)\frac{\vec\lambda \cdot \vec\sigma}{\|\vec\lambda\|}~.
\end{equation}
Using this equation with the fact that Pauli matrices are trace-less it is straightforward to obtain
\begin{align}
    P_{\text{save}} &= -\frac{2J^2 \tanh(\beta J \|\vec\lambda_A\|)}{\tau\|\vec\lambda_A\|}\int^\tau_0 dt \  \vec\lambda_t \cdot (\vec\lambda_B \wedge \vec\lambda_A )  ,\\
    C_{\text{save}}&=-\frac{4J^3}{\tau}\left[1 - \tanh(\beta J \|\vec\lambda_A\|)^2  \left(1- \frac{\vec\lambda_A \cdot\vec\lambda_B}{\|\vec\lambda_A\|^2}\right) \right] \int^\tau_0 dt \ \vec\lambda_t \cdot (\vec\lambda_B \wedge \vec\lambda_A ) ~.
\end{align}
It is clear that the integrands are maximised simultaneously by choosing a jump protocol with $\vec\xi =\vec\Lambda= \alpha \vec\lambda_A \wedge \vec\lambda_B$, and hence we find the optimal values
\begin{align}\label{eq:p_opt_q}
    P_{\text{save}}^*&=2\alpha J^2\sin^2\phi \|\vec\lambda_A\| \|\vec\lambda_B\|^2 \tanh(\beta J \|\vec\lambda_A\|)  ,\\
    \label{eq:c_opt_q}
    C^*_{\text{save}} &=4\alpha J^3\sin^2\phi \|\vec\lambda_A\|^2 \|\vec\lambda_B\|^2 \left[1 - \tanh(\beta J \|\vec\lambda_A\|)^2  \left(1- \frac{\|\vec\lambda_B\|}{\|\vec\lambda_A\|}\cos\phi\right) \right] ~,
\end{align}
where $\phi$ is the angle between $\vec\lambda_A$ and $\vec\lambda_B$. We can notice that $\tanh(\beta J \|\vec\lambda_A\|)^2  \left(1- \frac{\|\vec\lambda_B\|}{\|\vec\lambda_A\|}\cos\phi\right) < 1$ for all choices of $\vec\lambda_A$ and $\vec\lambda_B$. Therefore we can optimize fluctuations and excess work simultaneously by choosing $\alpha > 0$. The magnitude of $\alpha$ has to be chosen in such a way that that the error of the approximation~\eqref{eq:error_closed} remains small. A sufficient condition is then given by choosing 
\begin{equation}\label{eq:q_cond}
    0\leq\alpha \ll \frac{1}{J\tau |\sin\phi| \|\vec\lambda_A\| \|\vec\lambda_B\|}~.
\end{equation}
This condition is illustrated in Fig. \ref{fig:qubit}, in which we compare the results of \eqref{eq:p_opt_q} and \eqref{eq:c_opt_q} to the exact calculation of $P_{\text{save}}$ and $C_{\text{save}}$ for jump protocols in a qubit. The boundary conditions were set to $H_A = J\sigma^x$ ($\vec\lambda_A = \hat x$), $H_B = J\sigma^z$ ($\vec\lambda_B = \hat z$) and the relevant constants are set to $\tau J = \beta J = 1$. Then the condition of eq. \eqref{eq:q_cond} becomes $\alpha \ll 1$, indeed we can see from the figure that as $\alpha$ approaches $\mathcal{O}(1)$  the approximation breaks down. 

It is important to stress that for higher dimensional closed systems, simultaneous optimisation of $P_{\text{save}}$ and $C_{\text{save}}$ cannot always be guaranteed despite what we observe in the case of a qubit.

\section{Open quantum systems}\label{sec:classical}

\label{sec:open quantum systems}

\

We now move to open quantum systems. Our framework can be applied to any  Markovian Lindblad equation of the form~\eqref{eq:gen}, in which the generator $\mathcal{L}_{\vec{\lambda}_t}$ is independent of the velocity $d\vec{\lambda}_t/dt$ and depends only on $\vec{\lambda}_t$.  Here we  illustrate it for the simple evolution: 
\begin{align} \label{eq:markov_relaxation}
    \mathcal{L}_{\vec{\lambda}_t}[  \cdot  ]=\frac{\pi(\vec{\lambda})\tr{\cdot} -(\cdot)}{\tau^{eq}},
\end{align}
which describes a decay of the state $\rho$ into the instantaneous Gibbs state $\pi(\vec{\lambda})$ with a timescale $\tau^{eq}$. This dissipative evolution naturally arises in collisional models~\cite{Bumer2019} and also describes some systems weakly interacting with a reservoir with a sufficiently flat spectral density~\cite{Esposito2010}. For this type of dynamics we find some more illuminating expressions for the various terms appearing in the leading corrections to the excess work and variance. Firstly, the initial force relaxation rate becomes
\begin{align}
    \vec{R}(\vec{\lambda}):=\frac{\big\<\vec{X}\big\>_{\vec{\lambda}}-\big\<\vec{X}\big\>_A}{\tau^{eq}},
\end{align}
where  $\langle (.) \rangle_{\vec{\lambda}}=\tr{(.)\pi(\vec{\lambda})}$. This demonstrates that $\vec{R}(\vec{\lambda})$ quantifies the average rate at which each conjugate force changes from its initial value relative to the characteristic timescale $\tau^{eq}$. Furthermore, the correlation functions become
\begin{align}
    \big[\mathbf{G}(\vec{\lambda})\big]_{jk}=\frac{\mathcal{F}_{jk}(\vec{\lambda})-\mathcal{F}_{jk}(\vec{\lambda}_A)}{\tau^{eq}}+\bigg(\langle X_j \rangle_{\vec{\lambda}} \ R_k(\vec{\lambda})+\langle X_k \rangle_{\vec{\lambda}} \ R_j(\vec{\lambda})\bigg).
\end{align}
and 
\begin{align}
    \big[\mathbf{B}(\vec{\lambda})\big]_{jk}=-\frac{2}{\tau^{eq}}\mathcal{F}_{jk}(\vec{\lambda}),
\end{align}
where $\mathcal{F}_{jk}(\vec{\lambda})$ is the symmetric covariance defined as
\begin{align}\label{eq:metric}
    \mathcal{F}_{jk}(\vec{\lambda}):=\frac{1}{2}\tr{\{X_j,X_k\}\pi(\vec{\lambda})}-\langle X_j\rangle_{\vec{\lambda}}\langle X_k\rangle_{\vec{\lambda}}.
\end{align}
This function defines a metric tensor on the manifold of control parameters, and was first introduced in \cite{Unicersity} as a means of quantifying the geometric structure of thermal states. More recently this metric has been shown to determine optimal protocols with minimal work fluctuations in slow driving regimes, achieved by traversing a geodesic path in the parameter space~\cite{Miller2019,Miller2020}. In quasi-classical regimes where $[X_j,X_k]=0$ this metric becomes proportional to the well-known thermodynamic metric defined as the negative Hessian of the free energy \cite{Crooks2007}:
\begin{align}
    [X_j,X_k]=0 \mapsto \mathcal{F}_{jk}(\vec{\lambda})=-k_B T\frac{\partial^2 F}{\partial\lambda_j \partial\lambda_k} ,
\end{align}
This metric, which is equivalent to the fisher information matrix of the thermal state, can be used to determine geodesic paths with minimal excess work in slow driving and close to equilibrium regimes \cite{Sivak2012}. In the present context, we observe an interesting connection between the thermodynamic metric~\eqref{eq:metric} and the work fluctuations in the complete opposite regime of rapid driving. However, minimal fluctuations are not given by following a geodesic path, but rather they are achieved by jumping to a point in the parameter space that maximises the balance~\eqref{eq:max_c} between the change in metric components, relaxation rates $\vec{R}(\vec{\lambda})$ and displacements $\vec{\lambda}-\vec{\lambda}_B, \vec{\lambda}-\vec{\lambda}_A$. We will now explore the optimisation of three different  systems: a driven quantum dot, and two Ising spin chains.

\subsection{Fast erasure of a single bit}
\label{sec:erasureSingleQubit}

A driven quantum dot interacting weakly with an environment is a paradigmatic example of a system that can be described by the simple dynamics~\eqref{eq:markov_relaxation}~\cite{Esposito2010}. In that case the Hamiltonian is given by $H(\epsilon)=\frac{1}{2}\epsilon\sigma^z$ with a single control variable $\lambda_t=\epsilon(t)$ given by the energy gap of the two-level system. The optimal finite-time thermodynamics of such  systems has been well studied with regard to minimising average dissipation in Landauer erasure~\cite{Diana2013,Scandi2019,Zhen2021,VanVu2022,Zhen2022,Ma2022}, including a recent experimental implementation in a driven single dot~\cite{scandi2022constant}, as well as maximising  average power and efficiency  in heat engines~\cite{Esposito2010d,Cavina}. More recent numerical approaches have also been used to find optimal protocols that take into account the minimisation of  work fluctuations~\cite{Solon2018,Erdman2022}. In the present context, the fast driving assumption allows us to obtain analytic results for these optimal protocols, which we now apply to a rapid bit-erasure process.  The boundary conditions for erasure are then $\epsilon_A = 0$ and $\beta \epsilon_B \gg 1$. We first note that the leading term of the expansion \eqref{eq:av_expand} yields: 
\begin{equation}
    \frac{S(\pi(\vec{\lambda}_A)||\vec{\lambda}_B))}{\tau} \approx \frac{1}{\tau} \left(   \beta \epsilon_B - \ln 2 \right). 
\end{equation}
 To characterise the first order correction, straightforward calculations provide the following expressions:
\begin{align}
    \tau^{eq} R(\epsilon) &= \frac{1}{1+e^{\beta\epsilon}} - \frac{1}{2} ~,\\
    \tau^{eq} G(\epsilon) &= 0,\\
    \tau^{eq} B(\epsilon) &= -\frac{1}{2}~.
\end{align}
From this we can compute the power and constancy savings under the fast driving assumption $\tau/\tau_c\ll 1$:
\begin{align}
    P_{\text{save}}&= \frac{k_B T}{\tau^{eq}}\int_0^1 ds~ \big(\beta\epsilon_B-\beta\epsilon(s)\big)\bigg(\frac{1}{2} - \frac{1}{1+e^{\beta\epsilon(s)}} \bigg)  ,\\
  C_{\text{save}}&= \frac{k_B^2 T^2}{2\tau^{eq}} \int_0^1 ds~ \beta\epsilon(s)\big(\beta\epsilon_B-\beta\epsilon(s)\big).
\end{align}
We now seek to find the optimal energy gaps to jump to in order to maximise either $P_{\text{save}}$ or $C_{\text{save}}$. It will become clear in this case the power and constancy savings cannot be simultaneously maximised, and so the distinct gaps are denoted by $\xi$ and $\Lambda$ respectively. Maximising $P_{\text{save}}$ amounts to solving the following transcendental equation
 \begin{equation}
    \frac{1}{2}-\frac{1}{1+e^{\beta\xi}} = \frac{(\beta \epsilon_B - \beta\xi) e^{\beta \xi}}{(1+e^{\beta\xi})^2}~.
\end{equation}
In the limit of $\beta \epsilon_B \gg 1$ we can solve it analytically up to terms $\mathcal{O}(\beta^{-1}\epsilon_B^{-1}\ln{\beta\epsilon_B})$ and find the optimal jump $\epsilon\mapsto\xi = \beta^{-1}\ln{2\beta\epsilon_B}$. Maximum power savings are thus
\begin{align}
    P^*_{\text{save}} \simeq\frac{\epsilon_B}{2\tau^{eq}}.
\end{align}
For this power-optimised jump let us denote the resulting sub-optimal constancy savings by $C_{\text{save}}^\xi$:
\begin{align}
    C_{\text{save}}^\xi=\frac{\epsilon^2_B}{\tau^{eq}}  \frac{\ln{2\beta\epsilon_B}}{2\beta \epsilon_B}.
\end{align}
On the other hand, to maximise the constancy savings we need to choose a jump to $\epsilon \mapsto\Lambda = \epsilon_B/2$ instead. This yields
\begin{align}
     C_{\text{save}}^*=\frac{\epsilon_B^2}{8\tau^{eq}}  , \ \ \ \ \ \ \ \ \ \ \ P^\Lambda_{\text{save}} = \frac{\epsilon_B}{4\tau^{eq}}.
\end{align}
where the sub-optimal savings in power are denoted $P^\Lambda_{\text{save}}$. Clearly there exists a significant trade-off between these two choices of optimal protocol, with power-optimised jumps causing no improvement to the constancy while constancy-optimised jumps reducing the potential power savings by a factor of $1/2$. Further comparison can be made with that of a naive linear driving $\epsilon(t)= \epsilon_B t/\tau$, which results in savings given by 
\begin{align}
    P_{\text{naive}} \simeq\frac{\epsilon_B}{4\tau^{eq}}, \ \ \ \ \ \ \ \ \ \ \ C_{\text{naive}} =\frac{\epsilon_B^2}{12\tau^{eq}},
\end{align}
where we again drop terms of order $\mathcal{O}(\beta^{-1}\epsilon_B^{-1}\ln{\beta\epsilon_B})$. Therefore we can see that choosing an optimal jump for the excess work leads to an improvement factor of $1/2$, and choosing the optimal jump for the fluctuations gives an improvement factor of~$3/2$, each indicating significant improvements over a naive protocol. However, two unexpected observations here are that naive protocols are able to achieve larger savings in constancy than that of the power optimised protocol, and also achieve the same level of power savings to the constancy-optimised protocol. This emphasises that improvements to one objective do not necessarily translate into improvements of the other.

\subsection{Dissipative classical Ising chain}

\begin{figure}[H]
	\centering
	\includegraphics[width=\textwidth]{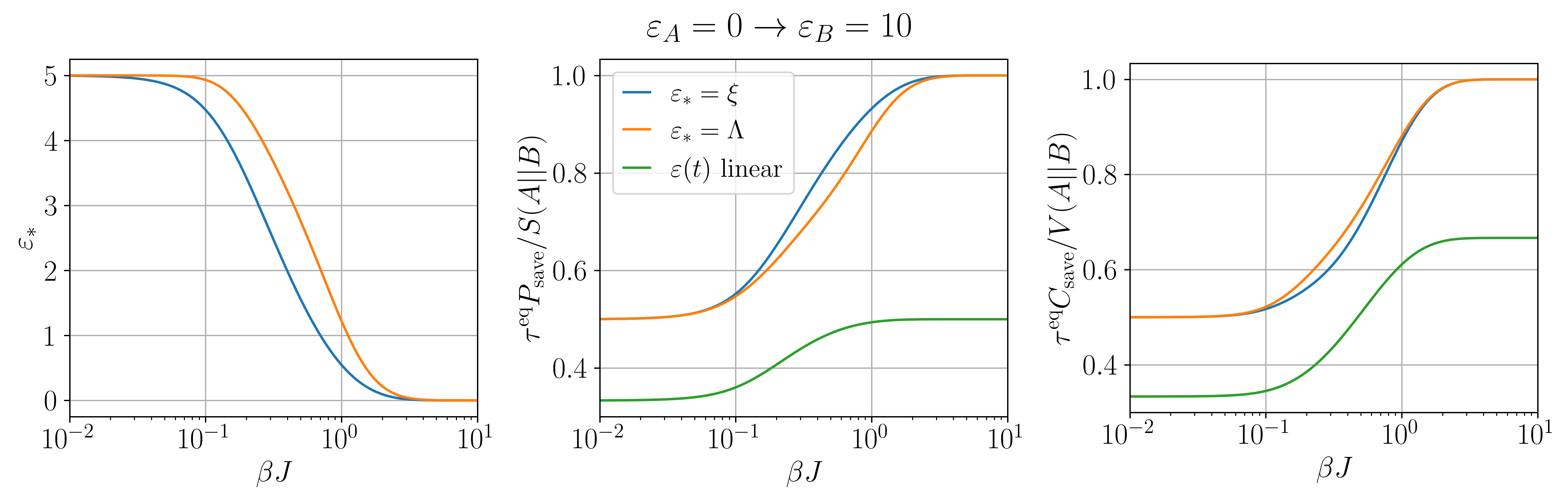}
	\caption{(left) Optimal value of $\varepsilon_*$ for a protocol going from $\varepsilon_A = 0$ to $\varepsilon_B=10$. (centre and right) Relative power and constancy savings in three protocols going from $\varepsilon_A = 0$ to $\varepsilon_B = 10$. We compare protocols that optimize excess work and fluctuations to a protocol that varies linearly the value of $\varepsilon$.}
	\label{fig:opt_Cising}
\end{figure}

The strength of our approach is that it enables to deal with more complex systems, where exact solutions for minimising dissipation and/or fluctuations are lacking -in contrast to the previous example in Sec. \eqref{sec:erasureSingleQubit}.  This is illustrated now for an Ising chain weakly coupled to a bath with dynamics~\eqref{eq:markov_relaxation}. We note that optimal driving protocols for classical spin chains have been devised in the slow driving regime~\cite{Rotskoff2015,Rotskoff2017,Louwerse2022}, and now we complement such results by studying the opposite fast-driving regime. We first consider a classical spin chain,
\begin{equation}
    H(\epsilon) = J\sum_{i=1}^n \left( \varepsilon\sigma^z_i - \sigma^z_i \sigma^z_{i+1} \right)~,
\end{equation}
where $J$ is the energy scale and $\varepsilon$ is a dimensionless parameter which can be interpreted as the strength of an external magnetic field. We assume that $\varepsilon$  can be controlled externally in time, $\lambda=\epsilon$, which results in work being done on the system. By assuming periodic boundary conditions we can compute the partition function per spin in the thermodynamic limit $n\to \infty$ \cite{Janyszek1989}:
\begin{equation}
    \lim_{n\rightarrow\infty}\frac{1}{n}\text{ln} \ Z = \beta J + \text{ln}\!\left[\cosh(\beta J\varepsilon) + \sqrt{\sinh(\beta J\varepsilon)^2 + e^{-4\beta J}}\right]~.
\end{equation}
The relevant conjugate force here is then the total $X= J\sum_i \sigma_i^z$. We now identify the following the relations
\begin{align}
    -\frac{1}{\beta}\frac{\partial}{\partial \varepsilon}\log Z &= \tr{X \pi(\varepsilon)}~,\\
    \frac{1}{\beta^2}\frac{\partial^2}{\partial \varepsilon^2}\log Z &= \tr{X^2 \pi(\varepsilon)} - \tr{X \pi(\varepsilon)}^2 ~.
\end{align}
These allow us to compute the first order corrections to the excess work and the fluctuations per site of the protocol in the thermodynamic limit from~\eqref{eq:fastwork} and~\eqref{eq:fast_var}, and employing the above expressions we obtain the initial force relaxation rates and correlation functions:
\begin{align}
    \tau^{eq} R(\varepsilon) &=  \frac{\sinh(\beta J \varepsilon_A)}{\sqrt{\sinh(\beta J\varepsilon_A)^2 + e^{-4\beta J}}} - \frac{\sinh(\beta J \varepsilon)}{\sqrt{\sinh(\beta J\varepsilon)^2 + e^{-4\beta J}}}~,\\
    \tau^{eq} G(\varepsilon) &= e^{-4\beta J}\left(\frac{\cosh(\beta J \varepsilon)}{(\sinh(\beta J\varepsilon)^2 + e^{-4\beta J})^{3/2}} - \frac{\cosh(\beta J \varepsilon_A)}{(\sinh(\beta J\varepsilon_A)^2 + e^{-4\beta J})^{3/2}}\right)  + (\tau^{eq} R(\varepsilon))^2 ,\\
    \tau^{eq} B(\varepsilon) &= -2 e^{-4\beta J} \frac{\cosh(\beta J \varepsilon_A)}{(\sinh(\beta J\varepsilon_A)^2 + e^{-4\beta J})^{3/2}}  ~.
\end{align}
It is now a case of substituting these into the two different Euler-Lagrange equations~\eqref{eq:EL_av} and~\eqref{eq:EL_var} to determine the optimal points $\xi$ and $\Lambda$ needed in each jump protocol, with solutions found numerically for a process that brings $\varepsilon$ from $\varepsilon_A=0$ to $\varepsilon_B=10$ (i.e. turning on the magnetic field). In Fig.~\ref{fig:opt_Cising} (left) we display the optimal field strength $\varepsilon_*=\{\xi,\Lambda\}$ that maximises either the power or constancy savings. We can notice that in the limits of high and low temperatures they coincide, while we cannot maximise them simultaneously in between these regimes. In Fig.~\ref{fig:opt_Cising} (centre) we plot the power savings $P^*_{\text{save}}$ relative to the zeroth contribution $k_B T S\big(\pi(\vec{\lambda}_A)||\pi(\vec{\lambda}_B)\big)/\tau_c$, while Fig.~\ref{fig:opt_Cising} (right) displays the  constancy savings $C^*_{\text{save}}$ in units of $k^2_B T^2 V\big(\pi(\vec{\lambda}_A)||\pi(\vec{\lambda}_B)\big)/\tau_c$. Both plots show the relative savings depending on whether we choose to optimise the power or constancy, and this is also compared to the savings achieved by taking a naive linear driving $\epsilon(t)=\epsilon_A(1-t/\tau)+\epsilon_B t/\tau$. In this case we can see that there is only a modest difference between the $\epsilon_A\mapsto\xi\mapsto \epsilon_B$ and $\epsilon_A\mapsto\Lambda\mapsto \epsilon_B$ jump protocols, and they each perform considerably better than the naive approach, contrasting with what we observed for the driven quantum dot. This highlights the importance of optimal control in many-body open quantum systems. 

\subsection{Ising chain in transverse field}\label{sec:quantum}

We will conclude with a final example covering the remaining scenario of an open quantum system where the control is such that the Hamiltonian may not commute at different times, so that $[H(\vec{\lambda}_t),H(\vec{\lambda}_{t'})]\neq 0$. This non-commutativity implies the presence of quantum friction \cite{Feldmann,Varizi2021a}, which is a distinctly non-classical contribution to the work done to drive the system that arises from allowing transitions between energy eigenstates. 
For this purpose we will consider a  dissipative Ising chain with simple dynamics~\eqref{eq:markov_relaxation}, though this time we apply a transverse field along the x-axis that can be controlled in time. We note that optimal driving protocols for this model has been considered in the slow driving regime~\cite{Scandi2019}, and the results presented here in the fast driving regime  are hence complementary. In particular, we will  focus on performing drivings close to a quantum phase transition, which has also been considered in several previous works~\cite{Polkovnikov2005,Dorner2012,Fusco2014a,Deffner2017}.

The Hamiltonian of the system is
\begin{equation}
    H(g) = -J\sum_{i=1}^n \left(\hat\sigma^z_i \hat\sigma^z_{i+1} + g\hat\sigma^x_i \right)~,
\end{equation}
where $J$ is the energy scale and $g$ is a dimensionless parameter which can be interpreted as an external (transverse) magnetic field. Clearly such a model will generate quantum friction as we vary the strength $g$ in time. Assuming again periodic boundary conditions we can compute the spectrum of the system with a Jordan-Wigner transformation \cite{sachdev_2011}. Then by taking the thermodynamic limit the partition function is given by
\begin{equation}
    \lim_{n\rightarrow\infty}\frac{1}{n}\log Z = \int_0^{2\pi}dk~ \log\!\left[2\cosh\frac{\beta\epsilon_k}{2}\right]~,
\end{equation}
where $\epsilon_k$ is the eigenenergy corresponding to the momentum $k$
\begin{equation}
    \epsilon_k = 2J\sqrt{1+g^2-2g\cos k}~.
\end{equation}
At zero temperature and $g=1$ this system presents a phase transition from an ordered ferromagnetic phase to a quantum paramagnetic phase. We will focus on studying protocols that take the system across this point by changing $g$ at finite temperature. The relevant conjugate force this time is $X = -J\sum_i \sigma_i^x$, and we can use the relations
\begin{align}
    -\frac{1}{\beta}\left.\frac{\partial}{\partial g}\log Z\right|_{g=g_*} &= \tr{ X \pi(g^*)}~,\\
    \frac{1}{\beta^2}\left.\frac{\partial^2}{\partial g^2}\log Z\right|_{g=g_*} &= \tr{ X^2 \pi(g^*)} - \tr{ X \pi(g^*)}^2. 
\end{align}
The first order corrections to the excess work and the fluctuations per site of the protocol are now computed within the thermodynamic limit, giving us
\begin{align}
    \tau^{eq} R(g_*) &= -\frac{1}{2}\left.\int_0^{2\pi}dk~ \dot\epsilon_k \tanh \frac{\beta\epsilon_k}{2}\right|^{g=g_*}_{g=g_A} ~,\\
    \tau^{eq} G(g_*) &= \frac{J}{2}\left.\int_0^{2\pi}dk~ \ddot\epsilon_k \tanh\frac{\beta\epsilon_k}{2} + \frac{\dot\epsilon_k^2}{2J} \cosh^{-2}\frac{\beta\epsilon_k}{2}\right|^{g=g_*}_{g=g_A} + (\tau^{eq} R(g_*))^2 ,\\
    \tau^{eq} B(g_*) &= -J\left.\int_0^{2\pi}dk~ \ddot\epsilon_k \tanh\frac{\beta\epsilon_k}{2} + \frac{\dot\epsilon_k^2}{2J} \cosh^{-2}\frac{\beta\epsilon_k}{2}\right|_{g=g_A} ~,
\end{align}
where $\dot \epsilon_k = \frac{d \epsilon_k}{dg}$. Substituting into the Euler-Lagrange equations~\eqref{eq:EL_av} and~\eqref{eq:EL_var} and solving numerically then determines the instantaneous jumps $g_A\mapsto\xi\mapsto g_B$ and $g_A\mapsto\Lambda\mapsto g_B$ for maximising the respective power and constancy savings. We set our boundary conditions to be $g_A=0$ and $g_B=3$ so that we turn on the transverse magnetic field and cross the phase transition point at $g=1$. Similarly to the classical case, we also compare these optimal protocols to a "naive" protocol in which the parameter is varied linearly in time, $g(t)=g_A(1-t/\tau)+g_B t/\tau$.

\begin{figure}[H]
	\centering
	\includegraphics[width=\textwidth]{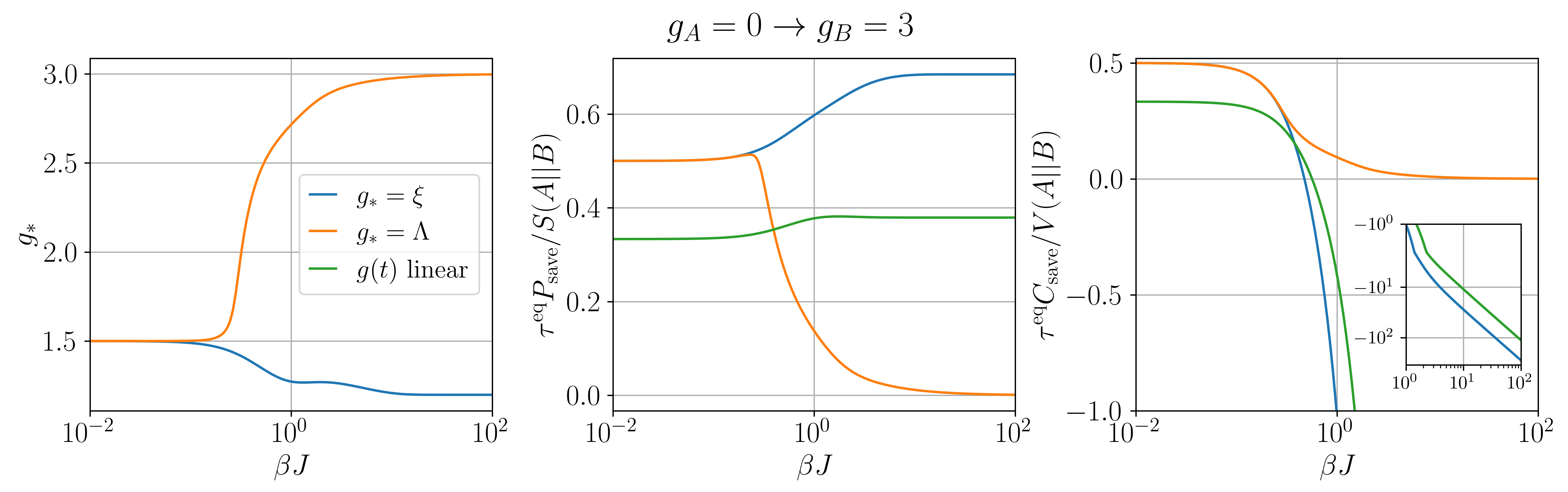}
	\caption{(left) Optimal value of $g_*=\{\xi,\Lambda\}$ for a protocol going from $g_A = 0$ to $g_B=3$. (centre) Relative power savings in three protocols going from $g_A = 0$ to $g_B = 3$ and (right) relative constancy savings for the same three protocols. In each figure we compare these optimal protocols to one that varies linearly the value of $g$.}
	\label{fig:opt_Qising}
\end{figure}

 In Fig.~\ref{fig:opt_Qising} (left) we display the optimal fields strength of $g_*=\{\xi,\Lambda\}$, which noticeably coincide in the limit of high temperatures like we saw with the classical Ising chain. On the other hand at low temperatures they no longer coincide, indicating a distinctly non-classical feature of this example and demonstrating that simultaneous optimisation is no longer possible. In Fig. \ref{fig:opt_Qising} we compare these two choices of protocol to linear driving and plot the resulting power savings (centre) and constancy savings (right). Since temperature is finite the phase transition is washed out but we can still observe a signature in the power and constancy savings occurring at lower temperatures, where we see that the two quantities move significantly further apart. One dramatic feature is the fact that optimising the power savings results in the constancy savings becoming significantly negative at lower temperatures beyond the phase transition, indicating a large growth in overall work fluctuations above that of an infinitely fast quench. On the other hand, if we choose to maximise the constancy savings we see that this drops drops to zero alongside the power savings at low temperatures. This indicates that the system is highly sensitive to the choice of protocol when driven close to a quantum phase transition.

\section{Discussion}

\

We have derived approximations for the average excess work done~\eqref{eq:fastwork} to rapidly drive a small system out of equilibrium along with the resulting work fluctuations~\eqref{eq:fast_var}. This has been derived under the assumption that $(i)$~the dynamics can be described by a Markovian generator that is independent of the velocities in the time-dependent control parameters, and $(ii)$~the duration of the process is short relative to the characteristic timescale of the dynamics, $\tau_c\gg \tau$. These approximations, which are based around the Dyson series, can be viewed as the inverse to the approximations used to treat the finite-time thermodynamics of slowly driven systems where $\tau_c\ll \tau$ \cite{Speck,Mandal2016a,Cavina2017,Miller2019}. Under these approximations we were able to prove that rapid processes that minimise either the average excess work or work fluctuations under fixed boundary conditions consist of two instantaneous jumps in the system control parameters, contrasting with the smooth geodesic paths that are optimal on slow driving and linear response regimes \cite{Sivak2012,Miller2019}, and bearing close similarity to exact results in stochastic thermodynamics~\cite{Schmiedl,Esposito2010}. Protocols that minimise the excess work done jump from the initial configuration to a point determined from the Euler-Lagrange equation~\eqref{eq:EL_av}, stay there for the duration $\tau$ then jump to the final boundary value. Protocols that minimise the work fluctuations follow the same pattern, but jump to an alternative point satisfying a different Euler-Lagrange equation~\eqref{eq:EL_var}. We have seen that in general, these points do no coincide which indicates a trade-off between the optimal values of the average and variance. While we have focused only on comparing these two distinct protocols, we can go further and reformulate this as a multi-objective optimisation problem. To do this one can use a \textit{Pareto front} to quantify the family of protocols interpolating between the minimal-dissipation and minimal-fluctuation processes \cite{Solon2018}. This is found from minimising the objective function $\alpha \langle W_{ex} \rangle+(1-\alpha)\sigma_W^2 \ \forall \alpha\in[0,1]$. By linearity, we can simply add~\eqref{eq:EL_av} to~\eqref{eq:EL_var} with respective weightings $\alpha$ and $1-\alpha$ to get the corresponding Euler-Lagrange equation. Crucially, Pareto-optimal solutions will again consist of instantaneous jumps in the control parameters. This extends the results of \cite{Blaber} to show that jump protocols continue to be optimal when one also cares about keeping fluctuations minimal, while also extending this approach to the full quantum regime. In particular, it is worth emphasising that, beyond the standard scenario of a driven system in contact with a Markovian environment, our approach also applies to closed quantum driven systems where the form of minimally dissipative driving processes remain less explored~\cite{Deffner,Sgroi2021}. Our work also complements  optimisations of power (and fluctuations) in quantum heat engines, where jump protocols were also typically found~\cite{Erdman2019,Cavina2020,Erdman2022}.   
Due to its generality, our optimisation scheme can be used to improve the control of complicated chemical, biological and quantum many-body systems whenever short operation times are desired. This has been illustrated by minimising both excess work and fluctuations for a classical and quantum spin chain where an external magnetic field is rapidly changed in time. When driving the system close to a quantum phase transition, we found  that optimising over driving protocols leads to substantial gains (see Fig. \ref{fig:opt_Qising}). 
It is interesting to combine and contrast these results with previous works considering the minimisation of dissipation in slowly driven many-body systems~\cite{Rotskoff2015,Rotskoff2017,Scandi2019,Louwerse2022}.

A number of improvements and generalisations to our approach are warranted. For open quantum systems, it is important to note that protocols with non-commuting Hamiltonians may not adequately be described by adiabatic Lindblad equations such as~\eqref{eq:markov_relaxation} when operating in the fast driving regime \cite{Erdman2019,Cavina2020}. Hence, 
one should find a way to match our fast driving approximation to an appropriate Markovian master equation in situations where the system is weakly-coupled to a bath while allowing for fast, non-commuting Hamiltonian protocols. 
Interestingly, since non-adiabatic corrections can potentially lead to a dependence on the control velocities \cite{Dann}, this would imply that instantaneous jumps are not necessarily optimal in these cases. The approximations~\eqref{eq:fastwork} and~\eqref{eq:fast_var} will remain valid in any case, but terms appearing inside the first order corrections will now depend on the rates $d\vec{\lambda}/dt$ and the Euler-Lagrange equations~\eqref{eq:EL_av} and~\eqref{eq:EL_var} will no longer hold. Moreover, this additional dependence means that solutions can generally consist of both discontinuous and continuous contributions to the optimal path, as is seen in exactly solvable models such as the driven Brownian particle \cite{Schmiedl}. Finding ways to determine these optimal protocols under more accurate treatments of the non-adiabatic dynamics will be crucial for understanding the impact of quantum fluctuations on the thermodynamics of rapidly driven systems.  As we saw with the toy model in Section~\ref{sec:quantum}, operating close to a quantum phase transition can have a significant detrimental impact on the work fluctuations along processes with minimal dissipation. For more realistic non-adiabatic dynamics, this motivates a more careful consideration of the full multi-objective optimisation problem in the presence of quantum fluctuations and will be left for future investigation. 

\emph{Acknowledgements. } A.R. and M. P.-L. acknowledge funding by the the Swiss National Science Foundation through an Ambizione Grant No. PZ00P2-186067. H. J. D. M. acknowledges support from the Royal Commission for the Exhibition of 1851.

\bibliographystyle{apsrev4-1}
\bibliography{mybib.bib}

\appendix

\section{Error approximation to fast driving expansion}

\

\noindent Choosing the jump protocol $\vec{\lambda}_t=\vec{\lambda}_A+[\vec{\zeta}-\vec{\lambda}_A]\theta(t)+[\vec{\lambda}_B-\vec{\zeta}]\theta(t-\tau)$ and using the fact that $\dot{\theta}(t-t')=\delta(t-t')$, we derive the following inequality:
\begin{align}\label{eq:approx_1}
    \nonumber |\langle W_{\text{ex}}\rangle^{\text{true}}-\langle W_{\text{ex}}\rangle^*|&=\bigg|\int^{1}_0 ds \ \frac{d\vec{\lambda}_s^T}{ds} \  \tr{\vec{X} \ \big(\tilde{\rho}(s)-\sigma(s)\big)}\bigg|, \\
    \nonumber&\leq \int^{1}_0 ds \ \bigg|\frac{d\vec{\lambda}_s^T}{ds}\tr{\vec{X} \ \big(\tilde{\rho}(s)-\sigma(s)\big)}\bigg|, \\
    \nonumber &\leq \int^{1}_0 ds \ ||\dot{H}(\vec{\lambda_s})||_1 ||\tilde{\rho}(s)-\sigma(s)\big)||_1, \\
    \nonumber&\leq\big(||H(\vec{\xi})-H(\vec{\lambda}_A)||_1+||H(\vec{\lambda}_B)-H(\vec{\xi})||_1\big)\mathcal{O}(\tau^2/\tau_c^2), \\
    \nonumber &\leq 2\max \big\{||H(\vec{\xi})-H(\vec{\lambda}_A)||_1,||H(\vec{\lambda}_B)-H(\vec{\xi})||_1\big\}\mathcal{O}(\tau^2/\tau_c^2), \\
    &= \Delta h(\vec{\xi}) \ \mathcal{O}(\tau^2/\tau_c^2)
\end{align}
where we used the triangle inequality followed by the trace bound $|\tr{AB}|\leq ||A||_1 ||B||_1$ and~\eqref{eq:approx}. A more involved procedure can be used to find the error approximation for the work fluctuations. First we define the truncated propagator appearing in the Dyson series~\eqref{eq:dyson},
\begin{align}
    \overleftarrow{M}(s,s')[(.)]:= (.) + \int^{s}_{s'} d\nu \ \tau\mathcal{L}_{\vec{\lambda}_\nu}[(.)],
\end{align}
For a given operator $A$ and $0\leq s\leq 1, \ 0\leq s'\leq 1$, consider the following norm
\begin{align}
    \mathcal{A}(s,s'):=\big|\big|\overleftarrow{P}(s,s')[&\Delta_{\tilde{\rho}(s')}A \  \tilde{\rho}(s')]-\overleftarrow{M}(s,s')[\Delta_{\sigma(s')}A \  \sigma(s')]\big|\big|_1
\end{align}
Expanding the propagators gives
\begin{align}
     \nonumber\mathcal{A}(s,s')&=\bigg|\bigg|\Delta_{\tilde{\rho}(s')}A \tilde{\rho}(s')-\Delta_{\sigma(s')}A \sigma(s')+\tau\int^{s}_{s'} d\nu \ \mathcal{L}_{\vec{\lambda}_\nu}[\Delta_{\tilde{\rho}(s')}A \tilde{\rho}(s')-\Delta_{\sigma(s')}A \sigma(s')] \\
     \nonumber& \ \ \ \ \ \ \ \ \ \ \ +\sum^{\infty}_{n=2}\tau^n\int^s_{s'} dt_{n} \int^{t_{n}}_{s'} dt_{n-1}...\int^{t_2}_{s'}dt_1 \  \mathcal{L}_{\vec{\lambda}_{t_{n}}}\mathcal{L}_{\vec{\lambda}_{t_{n-1}}}...\mathcal{L}_{\vec{\lambda}_{t_1}}[\Delta_{\tilde{\rho}(s')}A  \ \tilde{\rho}(s')]\bigg|\bigg|_1, \\
     &\leq \bigg|\bigg|\Delta_{\tilde{\rho}(s')}A \tilde{\rho}(s')-\Delta_{\sigma(s')}A \sigma(s')\bigg|\bigg|_1+\tau\bigg|\bigg|\int^{s}_{s'} d\nu \ \mathcal{L}_{\vec{\lambda}_\nu}[\Delta_{\tilde{\rho}(s')}A \tilde{\rho}(s')-\Delta_{\sigma(s')}A \ \sigma(s')]\bigg|\bigg|_1 \\
     \nonumber& \ \ \ \ \ \ \ \ \ \ \ +\bigg|\bigg|\sum^{\infty}_{n=2}\tau^n\int^s_{s'} dt_{n} \int^{t_{n}}_{s'} dt_{n-1}...\int^{t_2}_{s'}dt_1 \  \mathcal{L}_{\vec{\lambda}_{t_{n}}}\mathcal{L}_{\vec{\lambda}_{t_{n-1}}}...\mathcal{L}_{\vec{\lambda}_{t_1}}[\Delta_{\tilde{\rho}(s')}A  \ \tilde{\rho}(s')]\bigg|\bigg|_1 ,
\end{align}
where we used the triangle inequality for the trace norm. We next bound each of the three terms separately. Using the submultiplicavity of the trace norm, one has
\begin{align}
    \big|\big|\Delta_{\tilde{\rho}(s')}A \tilde{\rho}(s')-\Delta_{\sigma(s')}A \sigma(s')\big|\big|_1\leq 2\big|\big|A\big|\big|_1 \big|\big|\tilde{\rho}(s')-\sigma(s')\big|\big|_1=\big|\big|A\big|\big|_1\mathcal{O}(\tau^2/\tau_c^2) 
\end{align}
where we used~\eqref{eq:approx}. Similarly, using the definition of the characteristic timescale~\eqref{eq:timescale} one gets
\begin{align}
    \nonumber \tau\bigg|\bigg|\int^{s}_{s'} d\nu \ \mathcal{L}_{\vec{\lambda}_\nu}[\Delta_{\tilde{\rho}(s')}A \ \tilde{\rho}(s')-\Delta_{\sigma(s')}A \sigma(s')]\bigg|\bigg|_1 &\leq \frac{\tau(s-s')}{\tau_c} \big|\big| \Delta_{\tilde{\rho}(s')}A \tilde{\rho}(s')-\Delta_{\sigma(s')}A \sigma(s') \big|\big|_1, \\
    \nonumber &\leq \frac{2\tau}{\tau_c}  \big|\big|A\big|\big|_1 \big|\big|\tilde{\rho}(s')-\sigma(s')\big|\big|_1, \\
    &=2 \big|\big|A\big|\big|_1\mathcal{O}(\tau^3/\tau_c^3).
\end{align}
Furthermore,
\begin{align}
    \bigg|\bigg|\sum^{\infty}_{n=2}\tau^n\int^s_{s'} dt_{n} \int^{t_{n}}_{s'} dt_{n-1}...\int^{t_2}_{s'}dt_1 \  \mathcal{L}_{\vec{\lambda}_{t_{n}}}\mathcal{L}_{\vec{\lambda}_{t_{n-1}}}...\mathcal{L}_{\vec{\lambda}_{t_1}}[\Delta_{\tilde{\rho}(s')}A  \ \tilde{\rho}(s')]\bigg|\bigg|_1 \leq  \big|\big|A\big|\big|_1\mathcal{O}(\tau^2/\tau_c^2) ,
\end{align}
We can therefore conclude that 
\begin{align}\label{eq:boundA}
    \mathcal{A}(s,s')\leq \big|\big|A\big|\big|_1\mathcal{O}(\tau^2/\tau_c^2)
\end{align}
From here we can now bound the error in the fluctuations for the jump protocol $\vec{\lambda}_t=\vec{\lambda}_A+[\vec{\Lambda}-\vec{\lambda}_A]\theta(t)+[\vec{\lambda}_B-\vec{\lambda}]\theta(t-\tau)$, which we plug into
\begin{align}
     |(\sigma_W^2)^{\text{true}}-(\sigma^2_W)^*| &= 2 \left| \mathbb{R}e\!\int^1_0\! ds \int^s_0\! ds' \  \tr{\dot{H}(\vec{\lambda}_s)\left(\overleftarrow{P}(s,s')\big[\Delta_{\tilde{\rho}(s')}\dot{H}(\vec{\lambda}_{s'}) \tilde{\rho}(s')\big] - \overleftarrow{M}(s,s')\big[\Delta_{\sigma(s')}\dot{H}(\vec{\lambda}_{s'}) \sigma(s') \big] \right)}\right|
\end{align}
Following the same steps we applied in~\eqref{eq:approx_1} and combining this with~\eqref{eq:boundA}, it is straightforward to see that
\begin{align}
    |(\sigma_W^2)^{\text{true}}-(\sigma^2_W)^*|\leq \Delta h^2(\vec{\Lambda}) \ \mathcal{O}(\tau^2/\tau_c^2)
\end{align}

\end{document}